\documentclass[aps,prl,preprint,superscriptaddress]{revtex4}
\usepackage{graphicx}
\usepackage{float}
\usepackage{subfig}
\begin{document}

%\preprint{}

\title{Bound state energies using Phase integral methods}


\author{R.B. White}
\affiliation{Plasma Physics Laboratory, Princeton University, P.O. Box 451, Princeton, New Jersey 08543}

%\thanks{}
\author{A.G. Kutlin}
\affiliation{Institute of Applied Physics of Russian Academy of Sciences, 46 Ulyanov str., Nizhny Novgorod 603950, Russia}


\date{\today}

\begin{abstract}
The study of asymptotic properties of solutions to differential equations 
has a long and arduous history, with the most significant advances 
having been made in the development of quantum mechanics. A very 
powerful method of analysis is that of Phase Integrals, 
described by Heading. Key to this analysis are the Stokes constants 
and the rules for analytic continuation of an asymptotic 
solution through the complex plane.  These constants are easily determined 
for isolated singular points, by analytically 
continuing around them and, in the case of analytic functions, requiring 
the asymptotic solution to be single valued.  However, most interesting 
problems of mathematical physics involve several singular points. 
By examination of analytically tractable problems and more complex 
bound state problems involving multiple 
singular points, we show that the method of Phase Integrals can greatly 
improve the determination of bound state energy over the simple 
WKB values.  We also find from these examples that in the limit of 
large separation the Stokes constant for a first order singular 
point approaches the isolated singular point value. 

\end{abstract}

%\keywords{}

\maketitle

\section{introduction}
Many differential 
equations of interest can be put in the form
\begin{eqnarray}
\frac{d^2\psi}{dz^2} + Q(z,E)\psi = 0.   \label{wkbeqn}
\end{eqnarray}
For bound state problems the existence of solutions and the energy
eigenvalue  $E$ can often be determined by Phase Integral 
methods.
Briefly, the WKBJ (and often simply WKB) 
approximate solutions of Eq. \ref{wkbeqn}, so named after
Wentzel, Kramers, Brillouin, and Jeffreys\cite{wkbj}, take the form
\begin{eqnarray}
\psi_\pm = Q^{-1/4} e^{\pm i \int^z Q^{1/2} dz},   \label{wkbsol}
\end{eqnarray}
and provided that
$\left|\frac{dQ}{dz}Q^{-3/2}\right| \ll 1 $
a general solution of Eq. \ref{wkbeqn} can be approximated by
\begin{eqnarray}
\psi = a_+\psi_+ + a_-\psi_-.
\end{eqnarray}
The solutions $\psi_\pm$ are local, not global solutions of Eq. 
\ref{wkbeqn}. Clearly the
inequality is not valid in the vicinity of a zero of $Q(z,E)$, 
commonly called a turning point. Aside from this, however,  $\psi_\pm$ are not 
approximations of a continuous solution of Eq. \ref{wkbeqn} in the whole z 
plane. The method of Phase Integrals, described by  
Heading\cite{heading}, but first suggested by A. Zwann in his 
dissertation in 1929, consists in relating, for a given solution 
of Eq. \ref{wkbeqn}, the WKBJ approximation in one region of the $z$ plane to 
that in another\cite{rwbook,rwnum}. The solution then gives an approximation 
to the global solution of the given differential equation.  
 The WKB approximation has an extensive 
application in  quantum mechanics and other branches of physics, along 
with many attempts to improve its 
accuracy\cite{dunham,dingle73,berry90,berry91,sergeenko96,delabaere97,sergeenko02,mirnov10,poor16}.

The regions in the complex plane defining the validity of the individual 
approximations to the solution 
are separated by the so-called Stokes and anti-Stokes 
lines   associated with $Q(z,\omega)$, and thus the qualitative 
properties of the solution are determined once these lines are known. 
The global Stokes (anti-Stokes) lines associated with $Q(z,\omega)$ are paths in the 
$z$ plane, emanating from zeros or singularities of $Q(z,\omega)$, along which 
$\int Q^{1/2}(z,\omega)dz$ is imaginary (real).  When the zero $z_0$ 
is first order,
three anti-Stokes lines emanate from $z_0$.
Similarly, one finds that from a double root there issue four
anti-Stokes lines, from a simple pole a single line, etc..
 In refering to Stokes diagrams, we will refer to both 
zeros and singularities of Q(z) as singular points,  since it is the 
function $Q^{1/2}$ which is relevant in this diagram.

 Along the global anti-Stokes lines the functions 
$\psi_\pm$
are, within the validity of the WKBJ approximation, of constant
amplitude, $i.e.$ oscillatory.
 Along the Stokes lines the WKBJ solutions are exponentially
increasing or decreasing with fixed phase. Except at singular points, 
the Stokes and anti-Stokes lines are orthogonal. The global anti-Stokes
and Stokes lines which are attached to the singular points of the
Stokes diagram, along with the Riemann cut lines, determine the global
properties of the WKBJ solutions. 

In the notation of Heading, including the slow $Q^{-1/4}$ dependence, a 
WKBJ solution is denoted by
\begin{eqnarray}
(a,z)_s = Q^{-1/4}e^{i\int_a^z Q^{1/2} dz}   \label{eikon}
\end{eqnarray}
where the subscript s(d) indicates that the solution is
subdominant (dominant); 
$i.e.$ exponentially decreasing (increasing) for increasing
$|z-a|$ in a particular region of the $z$ plane, bounded by Stokes and
anti-Stokes lines.

Begin with a particular solution in one region of the $z$ plane, 
choosing that combination of subdominant and dominant solutions which 
gives the desired boundary conditions in this region. The global 
solution is obtained by continuing this solution through the whole $z$ 
plane effecting the following changes due to the monodromy of the 
WKB solutions\cite{poor16} moving through the plane:

1. Given a solution $\psi = a_d(z_0,z)_d + a_s(z,z_0)_s$, 
upon crossing a Stokes line emanating from $z_0$ in a
counterclockwise sense $a_s$ must be replaced by $a_s + Sa_d$ where S is
the Stokes constant associated with $z_0$.

2.	Upon crossing a cut in a counterclockwise sense, the cut
originating from a first order zero of $Q$ at the point $z_0$, we have 
\begin{eqnarray}
(z_0,z) \rightarrow -i(z,z_0)  \nonumber \\
(z,z_0) \rightarrow -i(z_0,z).
\end{eqnarray} 
Properties of dominancy or subdominancy are preserved in this 
process.

3.	Upon crossing an anti-Stokes line emanating from $z_0$, 
subdominant solutions attached to $z_0$ become 
dominant and vice versa.

4.  Reconnect from singularity $a$ to singularity $b$ using $(z,a) =
(z,b)[b,a]$ with $[b,a] = e^{i\int_b^a Q^{1/2} dz}$. 

Using these rules we can pass from region to region across the cuts,
Stokes and anti-Stokes lines emanating from the singularities. Beginning
with any combination of dominant and 
subdominant solutions in one region, this process leads to a globally
defined approximate solution of Eq. \ref{wkbeqn}.

For an isolated singular point with $Q$ analytic, the Stokes constant 
can be determined by a continuation around the complex plane, requiring 
that the WKB approximation be single valued. 
This gives for $Q = z^n$, the value $S = 2icos(\pi/(n+2))$. 

 Any function given on 
the real axis can be analytically continued into the complex plane, 
resulting in a collection of zeros and singularities associated with 
the functional form on the axis.  Every bound state problem consists 
of two turning points, with the potential positive outside them, and 
negative inside. The simplest bound state problem, the harmonic oscillator, 
given by the Weber equation with $Q = E-z^2$, is unique in that 
there are no additional zeros or singularities in the complex 
plane, the function is completely described by the zeros on the 
real axis. For this case the energy of the 
bound state as well as the single valuedness of the solution turn out 
to be independent of the value of the Stokes constant but the equation 
can be solved analytically, giving the Stokes constant for any energy. 

In section \ref{web} we review this case.  In section \ref{budden} we 
examine the Budden problem, another case with two singular points which 
admits an analytic solution.   In sections 
\ref{sz4} and \ref{sz6} we consider Hermitian anharmonic oscillator 
Hamiltonians with additional 
zeros above and below the axis, and in section \ref{sz3} we consider 
a non-Hermitian Hamiltonian with three singular points. The shape of these 
potentials is shown is section \ref{pot}.  Finally in 
section \ref{con} are the conclusions.

\section{The Weber function \label{web}}

\begin{figure}
\centering
\noindent
\includegraphics[scale=.5]{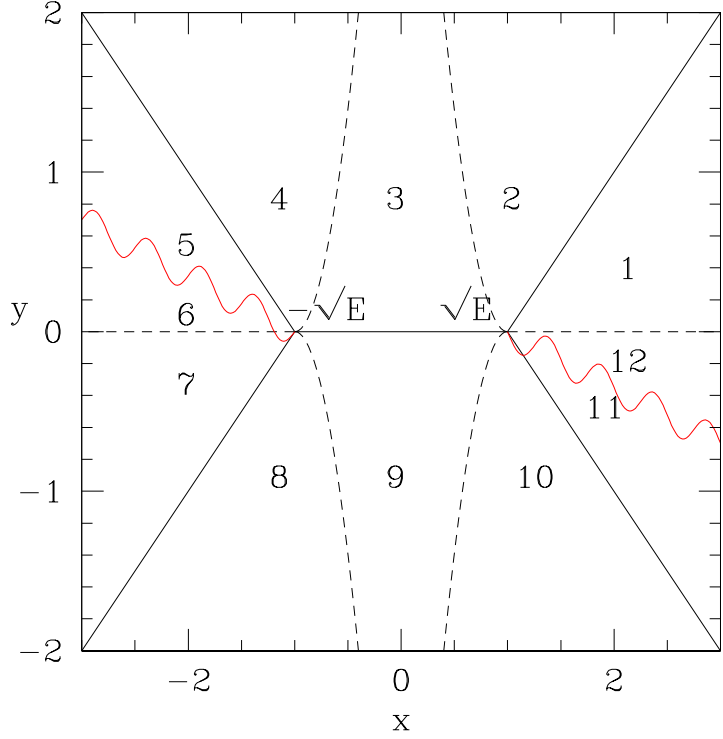}
\caption{Stokes plot for the bound state problem $Q = E - z^2$.}
\label{weberf}
\end{figure}

The simplest bound state problem is given by the harmonic
 oscillator potential, 
with $Q(z) = E - z^2$, real on 
the real axis with two first order zeros at points $\pm \sqrt{E}$.
 The Stokes diagram
is shown in Fig. \ref{weberf}.  The solid lines are Anti Stokes lines, 
dashed lines are Stokes lines, and cuts are designated with a wavy line. It is important to note that the results of any calcuation are 
independent of the location of the Riemann cuts, so they can be placed 
to be convenient for a given calculation.  
Denoting the Stokes
constant as $S$,  we will find that the boundary conditions immediately give
the Bohr--Sommerfeld  condition, which determines
the energy of the bound state, independent of the value of the Stokes constant. 

Begin with a subdominant solution at large positive $x$ and continue, 
assuming Stokes constants the same at each vertex, with $a = \sqrt{E}$:\\
(1)  $(a,z)_s $\\
(2)  $(a,z)_d $\\
(3)  $(a,z)_d + S(z,a)_s $ \\
(3)  $[a,-a](-a,z)_d + S(z,-a)_s[-a,a] $ \\
But $[-a,a] = e^{-iW}$, $W = \int_{-a}^a\sqrt{a^2-x^2}dx$. Note that the 
phase of $\sqrt{Q}$ is defined by requiring that $(a,z)_s$ be subdomanant 
in domain 1, so $i\sqrt{Q} = - real$.  The cut locations define the phase 
in other domains. \\
(4) $e^{iW}(-a,z)_d + S(z,-a)_s(e^{-iW} + e^{iW}$)\\
(5) $e^{iW}(-a,z)_s + S(z,-a)_d 2cosW $\\ 
(6) $-ie^{iW}(z,-a)_s -i S(-a,z)_d 2cosW $\\ 
Now set the dominant term to zero giving $W = (n + 1/2)\pi$, 
the Bohr Sommerfeld condition, independent of the value of the Stokes constant.
Note that $-ie^{iW} = (-1)^n$, the coefficient correctly reflects even and 
odd symmetry of the solution.
Continuing around to domain (12)\\ 
(7) $-ie^{iW}(z,-a)_s $\\
(8) $ -ie^{iW}(z,-a)_d $\\
(9) $-ie^{iW}[(z,-a)_d + S(-a,z)_s] $ \\
(9) $-i e^{iW}[(z,a)_d[a,-a]  +S [-a,a](a,z)_s]$ \\
(9) $-ie^{2iW}(z,a)_d - iS(a,z)_s $ \\
(10) $i(z,a)_d  $\\
(11) $i(z,a)_s  $\\
(12) $ (a,z)_s$,\\ 
we find again a subdominant solution with
 coefficient equal to one, so the solution is single valued, independent 
of the value of S provided that 
the Bohr-Sommerfeld condition on the energy is satisfied.  
Of course every Stokes constant 
has a definite value, which can be revealed from an exact solution.

A general solution of the Weber equation can be written with use of 
parabolic cylinder functions: 
\begin{eqnarray}
 y(z)=C_1 U(-E/2,\sqrt{2}z)+C_2 U(E/2,i\sqrt{2}z).
\label{WebGenSol}
\end{eqnarray}
To find Stokes constants in the upper half-plane of the complex z-plane 
let us choose $C_1$ and $C_2$ such that $y(z) \sim (a,z)$ for 
$-\pi/4<Arg(z)<\pi/4$. According to Eq. \ref{eikon}
\begin{eqnarray}
 (a,z)=(E-z^2)^{-1/4}e^{i\int_a^z (E-z^2)^{1/2} dz}=(E-z^2)^{-1/4}e^{-z^2/2 + E/4 + (E/2) \ln{(2z/\sqrt{E})}+O(1/z)},
\end{eqnarray}
and in a limit of large $z$ we can write
\begin{eqnarray}
 (a,z) \sim e^{i\pi/4}\left(2\sqrt{\frac{e}{E}}\right)^{E/2}
z^{-1/2+E/2}e^{-z^2/2},\\
(z,a) \sim e^{i\pi/4}\left(2\sqrt{\frac{e}{E}}\right)^{-E/2}z^{-1/2-E/2}e^{z^2/2}.
\end{eqnarray}
The common factor $e^{i\pi/4}$ can be omitted for further calculations 
due to the linearity of the Weber equation.

Using asymptotic expansions for the parabolic cylinder functions 
we find that the required values for the arbitrary constants are
\begin{eqnarray}
 C_1=2^{-1/4}(2e/E)^{-E/4}, C_2=0.
\end{eqnarray}
Once we have determined the values of the arbitrary constants in 
Eq.\ref{WebGenSol}, we can write an asymptotic expansion of our 
solution in any domain. Particularly, for $\pi/4<Arg(z)<3\pi/4$ it has 
the form
\begin{eqnarray}
y(x) \sim (a,z)_d+i\frac{\sqrt{2\pi}e^{i\pi E/2}}{\Gamma(\frac{1-E}{2})}(2e/E)^{E/2}(z,a)_s.
\end{eqnarray}
Now compare this relation with the one from the WKB analysis performed 
above, line $4$. One might think that the coefficient of the subdominant 
function is equal to $S(e^{-iW}+e^{iW})$ but this is not quite true. 
Since Stokes constants depend on the lower limit of integration used in 
the WKB-approximation we should at first write everything in a unique 
base and only then compare rigorous result with the WKB one. 
And since $(a,z)=[a,-a](-a,z)$ and $(z,a)=(z,-a)[-a,a]$ we can write 
\begin{eqnarray}
S(1+e^{2iW})=i\frac{\sqrt{2\pi}e^{i\pi E/2}}{\Gamma(\frac{1-E}{2})}(2e/E)^{E/2}.
\end{eqnarray}
As long as we chose a branch of the square root such that $(a,z)$ is 
subdominant for large positive z, $W=\pi E/2$ and
\begin{eqnarray}
S=i\frac{\sqrt{2\pi}e^{i\pi E/2}}{\Gamma(\frac{1-E}{2})}(2e/E)^{E/2}(1+e^{i\pi E})^{-1}.
\end{eqnarray}
Now one can easily see that $S$ approaches $i$ for large energies.

\begin{figure}
\centering
\noindent
\includegraphics[scale=.8]{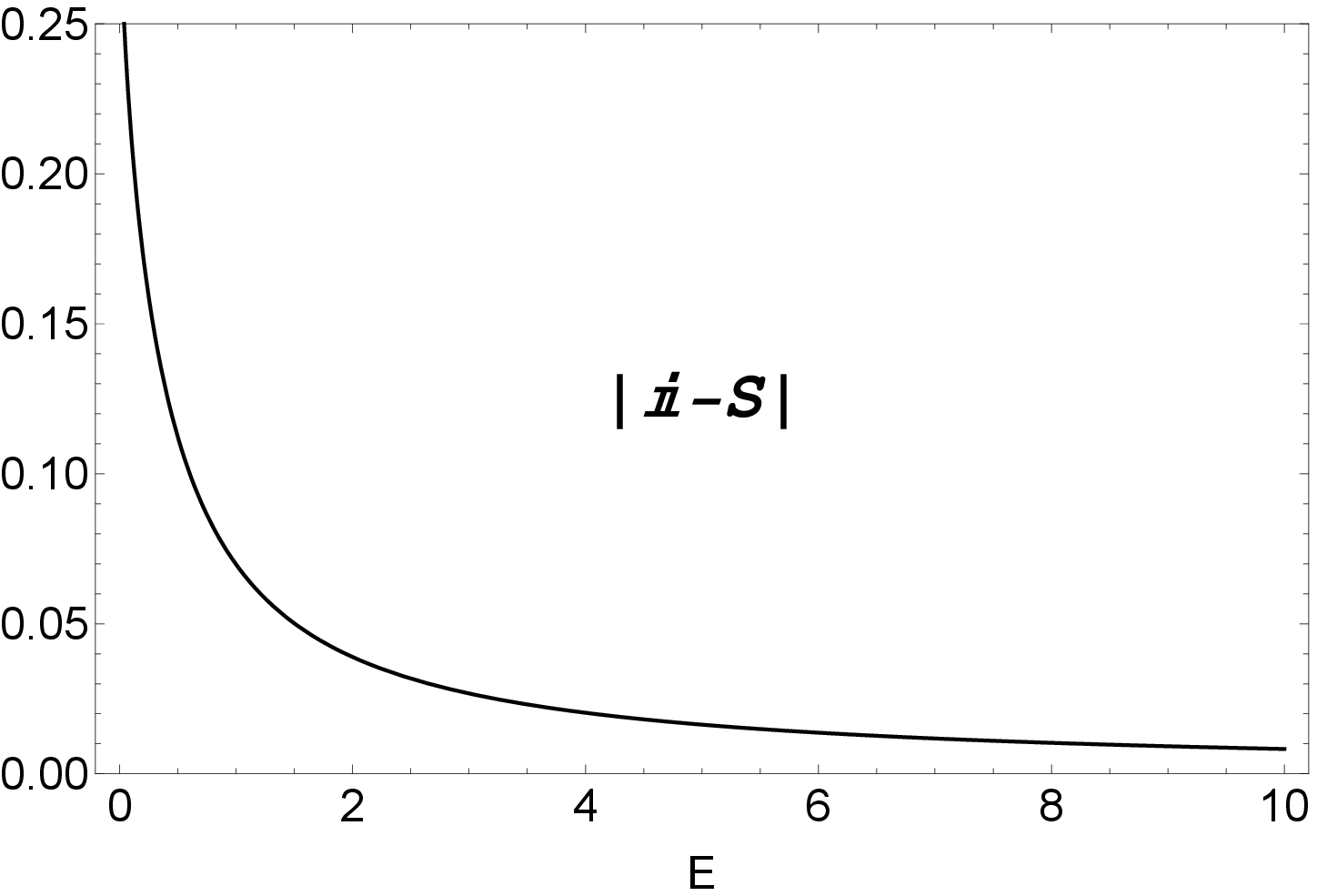}
\caption{Thedifference between $i$ and the actual value of the Stokes 
constant for the Weber equation}
\label{WebDiff}
\end{figure} 

\newpage

\section{The Budden equation \label{budden}}

Another example of a potential which allows exact calculation 
of Stokes constants is the Budden potential with $Q(z)=1+c/z$. 
We will see that the Stokes constants approach $i$ for
 large values of $c$ exactly as in the previous chapter.

\begin{figure}
\centering
\noindent
\includegraphics[scale=.5]{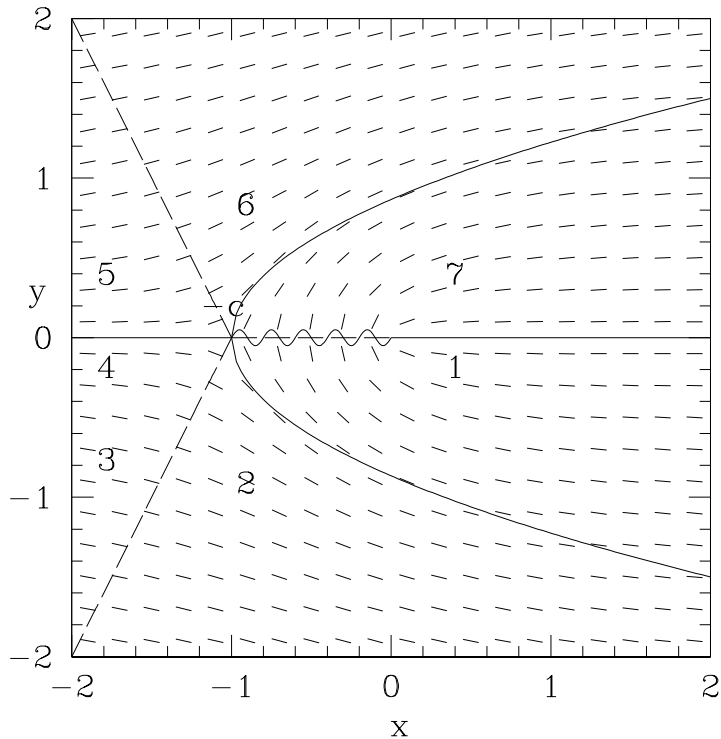}
\caption{Stokes plot for the Budden problem $Q = 1 + c/z$.}
\label{budd}
\end{figure} 

Before we calculate Stokes constants let us perform a WKB 
analysis. Choosing zero as a lower limit of integration we have
\begin{eqnarray}
i \int_0^z \sqrt{1+\frac{c}{x}} dx = i z + \frac{i c}{2} 
- \frac{i c}{2} \ln{\left(\frac{c}{4 z}\right)}+O\left(\frac{1}{\sqrt{z}}\right)
\end{eqnarray}
and in a limit of large z
\begin{eqnarray}
(0,z) \sim \left(\frac{4e}{c}\right)^{ic/2} z^{ic/2} e^{i z},\\
(z,0) \sim \left(\frac{4e}{c}\right)^{-ic/2} z^{-ic/2} e^{-i z}.
\label{BudBase}
\end{eqnarray}
To perform an analytical continuation around the origin, place the 
cut between \mbox{$z=0$} and \mbox{$z=-c$} along the real axis as shown 
in Fig.\ref{budd}. Starting with \mbox{$y(z) \sim (0,z)$} at large 
negative $z$ and using the rules of a  continuation, one finds that
 the continuation is \\
$3.\ (0,z)_d = [0,-c](-c,z)_d$\\
$2.\ [0,-c](-c,z)_d + S [0,-c] (z,-c)_s$\\
$1.\ [0,-c](-c,z)_s + S [0,-c] (z,-c)_d = (0,z)_d + S [0,-c]^2 (z,0)_s$\\
All integrals here were evaluated below the cut and give 
$[0,-c]=e^{\frac{\pi c}{2}}$.

\begin{figure}
\centering
\noindent
\includegraphics[scale=.8]{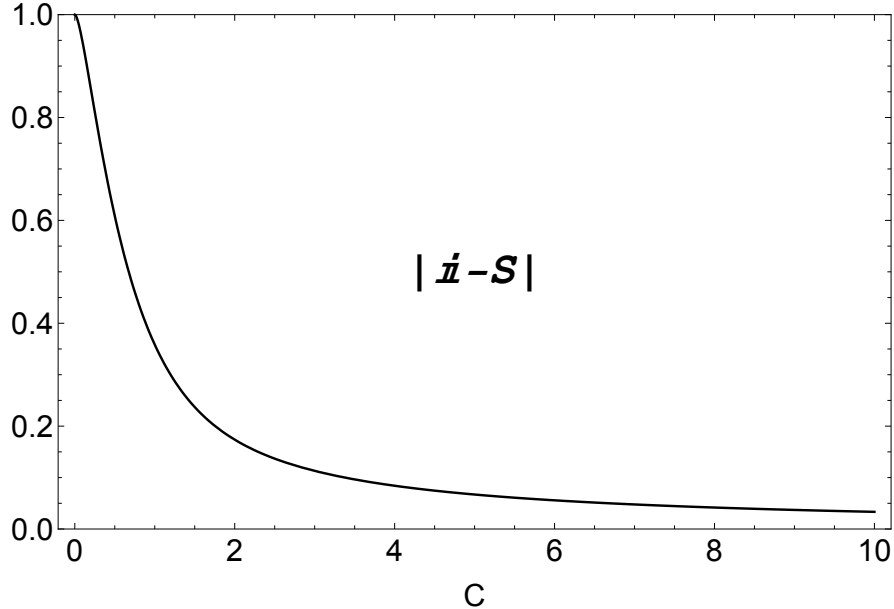}
\caption{The difference between $i$ and actual value of Stokes constant 
for the Budden equation}
\label{BudDiff}
\end{figure}

Now we can find an exact value of $S$. A general solution for the 
Budden equation is
\begin{eqnarray}
 y(z)=z e^{-iz}[C_1 U(1+\frac{ic}{2},2,2iz)+C_2 M(1+ic/2,2,2iz)],
\label{BudGenSol}
\end{eqnarray}
where $U(a,b,z)$ and $M(a,b,z)$ are Kummer functions. Requiring the 
solution $y(z)$ to be asymptotic to $(0,z)$ for large negative $z$ we find that
\begin{eqnarray}
 C_1 = \frac{\Gamma(1 + \frac{ic}{2})}{\Gamma(1 - \frac{ic}{2})} 
e^{3 \pi c/4} \left(\frac{2e}{c}\right)^{ic/2},\\
C_2 = \Gamma(1 + \frac{ic}{2}) e^{\pi c/4} \left(\frac{2e}{c}\right)^{ic/2}.
\end{eqnarray}
Now, using an asymptotic expansion of this solution for large positive z
 we have
\begin{eqnarray}
 y(z) \sim (0,z) + \frac{\Gamma(1 + \frac{ic}{2})}{\Gamma(1 - \frac{ic}{2})} 
\left(\frac{2e}{c}\right)^{ic} (e^{\pi c} - 1)(z,0).
\end{eqnarray}
Finally, comparing this asymptotic relation with the one from our WKB analysis we deduce that
\begin{eqnarray}
 S = \frac{\Gamma(1 + \frac{ic}{2})}{\Gamma(1 - \frac{ic}{2})} (\frac{2e}{c})^{ic} (1 - e^{-\pi c}).
\end{eqnarray}
It can be easily verified that this Stokes constant approaches $i$ for 
large values of $c$, as shown in Fig. \ref{BudDiff}.

\newpage

\section{A fourth order potential \label{sz4}}

\begin{figure}
\centering
\noindent
\includegraphics[scale=.5]{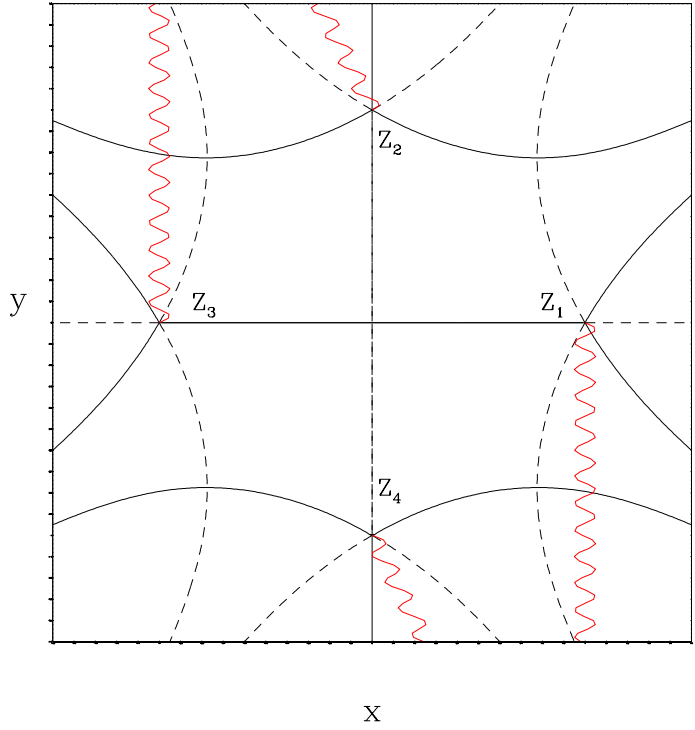}
\caption{Stokes plot for the bound state problem $Q = E - z^4$.}
\label{z4}
\end{figure}

Now consider a more complicated potential, that of an anharmonic oscillator,
 with more turning points in the complex plane. 
 As an example we take  $Q(z) = E - z^4$.  The Stokes structure is 
shown in Fig. \ref{z4}.  
The cuts have been chosen to give symmetry in the 
continuation between the upper and lower half planes.  We assume the 
Stokes constants to be the same at all singular points.  This is probably 
not true for general separation, but we will be restricted to an asymptotic 
evaluation for large separation, and this turns out to be true to leading 
order. 

In order to do the connections, we need the expressions 
$[k,l] = e^{\int_{z_k}^{z_l}i\sqrt{Q(z)}dz}$.  Note that the 
sheet of $i\sqrt{Q(z)}$ is defined by the cut locations, with 
the initial sheet determined by the fact that $(z_1,z)$ is 
subdominant for $x \rightarrow +\infty$ , meaning that
 $i\sqrt{Q}(z) = -real$ in this domain.

Carrying out the integrals then gives 
\begin{eqnarray}
[1,2] = e^{W/2}e^{iW/2},\hspace{5mm}  [1,3] = e^{iW},\hspace{5mm} 
 [2,3] = e^{-W/2}e^{iW/2}
\end{eqnarray}
where $W = E^{3/4}\int_{-1}^1\sqrt{1 - u^4}du$,  
and $\int_{-1}^1\sqrt{1 - u^4}du = 1.74804$.

Begin with a subdominant solution $\psi(z) = (Z_1,z)_s$
 at large positive $x$ and
continue through the upper half plane above the singularity at 
$iE^{1/4}$ to large negative $x$.
Using the symmetry of the potential we have 
\begin{eqnarray}
 Se^{-W} + Se^{iW} = -(-1)^n.
\label{mRz4}
\end{eqnarray} 
Also we find a condition for the vanishing of the 
dominant solution 
\begin{eqnarray}
e^{W}[1 + S^2] + 2S^2 cos(W) + e^{-W}S^2 = 0.
\label{subz4}
\end{eqnarray}

For large turning point separation $W$ is large and we have a solution given   
by the isolated turning point values, $S = i$ 
and $W_n = (n+1/2)\pi$, the usual approximate WKB solution.
There is a natural perturbation expansion parameter given by the existence 
of the exponential term $e^{-W}$, present because of the additional 
singular points not on the real axis. 
Even for the lowest bound state as given by the WKB approximation 
  $e^{-W_0} \simeq 0.2$, and for the next level
 $e^{-W_1} \simeq 0.009$.

Perturbing about the WKB value $W_n = (n+1/2)\pi$ gives
 the solution  
\begin{eqnarray}
S \simeq i\left[1 - cos(W_n)e^{-W_n} - \frac{e^{-2W_n}}{2}\right],
 \hspace{1cm} cos(W) = -e^{-W_n}.
\label{sval4}
\end{eqnarray}
 Thus $S \simeq i(1 + e^{-2W}/2)$ and 
 $sin(W) = (-1)^n\sqrt{1 - cos^2(W)} \simeq (-1)^n(1 - e^{-2W}/2)$, 
where we have dropped terms of order $e^{-4W}$, amounting to a correction 
to the ground state energy of a fraction of a percent.  

Values of the exact energy levels, the WKB approximation, and the 
Phase Integral evaluation are shown in table I.
For the ground state $E_{PI}$ has a 4 percent error, 
$E_{WKB}$  has a 20 percent error. 

This problem has also been approached using higher order phase integral
 approximations\cite{campbell,bender77}.  A third 
order solution using the phase integral series due to Froman\cite{froman} is given in table 2.1 
of Child\cite{child}.  The third order approximation to the ground state eigenvalue is 
given as 0.98076, with an error over 7 percent.

\begin{eqnarray}
{\bf Table \hspace{1mm} I. \hspace{2mm} 
Energy \hspace{2mm}  Levels  \hspace{2mm}  Q = E - z^4 }\nonumber \hspace{2mm}\\
\nonumber
\left( \begin{array}{c|c|c|c|c}
\hline
n & E_{exact}  & E_{wkb} & cos(W) & E_{PI} \\
\hline
0 & 1.0604 & 0.8671&   -0.207879 & 1.0246 \\
1 & 3.7964 & 3.7519 &   -8.9833\times 10^{-3} & 3.7424 \\
2 & 7.45567 & 7.4139 &  -3.8820\times 10^{-4} & 7.4144 \\
3 & 11.6374 & 11.6114 &  - 1.6776\times 10^{-5} & 11.6114 \\
4 & 16.2618 & 16.2335 &  - 7.2495\times 10^{-7} & 16.2335 
\nonumber
\end{array}\right)
\label{ez4}
\end{eqnarray}

\newpage

\section{A sixth order potential \label{sz6}}

\begin{figure}
\centering
\noindent
\includegraphics[scale=.5]{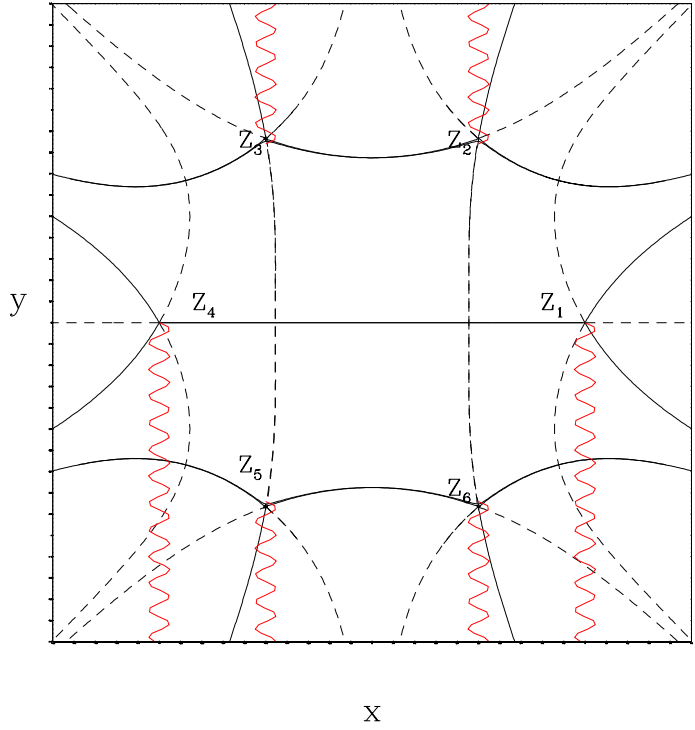}
\caption{Stokes plot for the bound state problem $Q = E - z^6$.}
\label{z6}
\end{figure}

Now consider an anharmonic oscillator 
 with more singular points in the complex plane,  $Q(z) = E - z^6$.
  The Stokes structure is 
shown  in Fig. \ref{z6} with first order zeros located at
 $Z_k = E^{1/6}e^{i(k-1)\pi/3}$ with $k = 1,2,3,4,5,6$.   We assume the 
Stokes constants to be the same at all singular points.

In order to do the connections, we need the expressions 
$[k,l] = e^{\int_{z_k}^{z_l}i\sqrt{Q(z)}dz}$.  Note that the 
sheet of $i\sqrt{Q(z)}$ is defined by the cut locations, with 
the initial sheet determined by the fact that $(z_1,z)$ is 
subdominant for $x \rightarrow +\infty$ , meaning that
 $i\sqrt{Q}(z) = -real$ in this domain.
Carrying out the integrals then gives 
\begin{eqnarray}
[1,2] = e^{\sqrt{3}W/4}e^{iW/4},\hspace{5mm}  [3,2] = e^{-iW/2},\hspace{5mm} 
 [3,4] = e^{-\sqrt{3}W/4}e^{iW/4},
\end{eqnarray}
where $W = E^{2/3}\int_{-1}^1\sqrt{1 - u^6}du$, 
and $\int_{-1}^1\sqrt{1 - u^6}du = 1.821488$.

Begin with a subdominant solution $\psi(z) = (Z_1,z)_s$
 at large positive $x$ and
continue through the upper half plane above all singularities 
to large negative $x$.
Choosing the solution to be real for $x \rightarrow \infty$ and using
the symmetry of the potential, but also noting that with the choice of cuts 
we have $Q^{1/4} =  e^{i\pi/2}$ for  large positive $x$ and 
 $Q^{1/4} =  e^{-i\pi/2}$ for  large negative $x$  we find
\begin{eqnarray}
(-1)^ni = 1 + S^2[1 + cos(W) + isin(W) + 2e^{-\sqrt{3}W/2}cos(W/2)].
\label{mRz6}
\end{eqnarray}
Also we find a condition for the vanishing of the 
dominant solution 
\begin{eqnarray}
1 + e^{\sqrt{3}W/2}cos(W/2) 
+ S^2[e^{\sqrt{3}W/2} + 2cos(W/2) + e^{-\sqrt{3}W/2}]cos(W/2) = 0,
\label{subz6}
\end{eqnarray}
giving $S^2 = -1 + O(e^{-\sqrt{3}W})$.

For large turning point separation $W$ is large and we have a solution given   
by the isolated turning point values, $S = i$ 
and $W_n = (n+1/2)\pi$, the usual approximate WKB solution.
A first order perturbation about the WKB value $W_n$ gives
 the solution  
\begin{eqnarray}
S = i, \hspace{1cm} cos(W) = -2e^{-\sqrt{3}W_n/2}cos(W_n/2).
\label{sval6}
\end{eqnarray}
It is interesting to note that these solutions break down at the second 
order in $\epsilon = e^{-\sqrt{3}W/2}$, meaning that one or more 
of the Stokes constants has a second 
order correction not given by Eq. \ref{subz6}.  Note that Eq. \ref{subz6} 
is real, but expanding $sin(W)$ in Eq. \ref{mRz6} through 
$sin(W) \simeq (-1)^n(1 - cos^2(W)/2)$ gives an additional second order 
imaginary term, but there is no second order term to balance it, and 
we conclude that $S^2$ 
must possess an imaginary second order term, not given by Eq. \ref{subz6}, 
so this equation can be trusted only to first order.

Values of the exact energy levels, the WKB approximation, and the 
Phase Integral evaluation are shown in table II.
For the ground state $E_{PI}$ has a 4 percent error, 
$E_{WKB}$  has a 30 percent error. 

\begin{eqnarray}
{\bf Table \hspace{1mm} II. \hspace{2mm} 
Energy \hspace{2mm}  Levels  \hspace{2mm}  Q = E - z^6 }\nonumber \hspace{2mm}\\
\nonumber
\left( \begin{array}{c|c|c|c|c}
\hline
n & E_{exact}  & E_{wkb} & cos(W) & E_{PI} \\
\hline
0 & 1.1448 & 0.8008&   -0.36206 & 1.1009 \\
1 & 4.3332 & 4.1612 &   2.3888\times 10^{-2} & 4.1929 \\
2 & 9.0731 & 8.9535 &   1.5727\times 10^{-3} & 8.9508 \\
3 & 14.9195 & 14.8316 &  -1.0354\times 10^{-4} & 14.8314 \\
4 & 21.7140 & 21.6224 &  -6.81617\times 10^{-6} & 21.6224 
\nonumber
\end{array}\right)
\label{ez6}
\end{eqnarray}

\section{ A non hermitian hamiltonian \label{sz3}}

Non-Hermitian Hamiltonians having $PT$ symnmetry have been 
shown to have real spectra, following a conjecture by D. Bessis  
that the spectrum of the Hamiltonian $H = p^2 + x^2 + ix^3$ is real 
and positive.  
A non-Hermitian Hamiltonian problem studied by 
Bender and Boettcher\cite{bender} 
is given by the function $Q(z) = E + (iz)^N$.  The energy spectrum is 
positive because of symmetry under the product of parity and time 
reversal.  As an example we take $N = 3$. 

\begin{figure}
\centering
\noindent
\includegraphics[scale=.5]{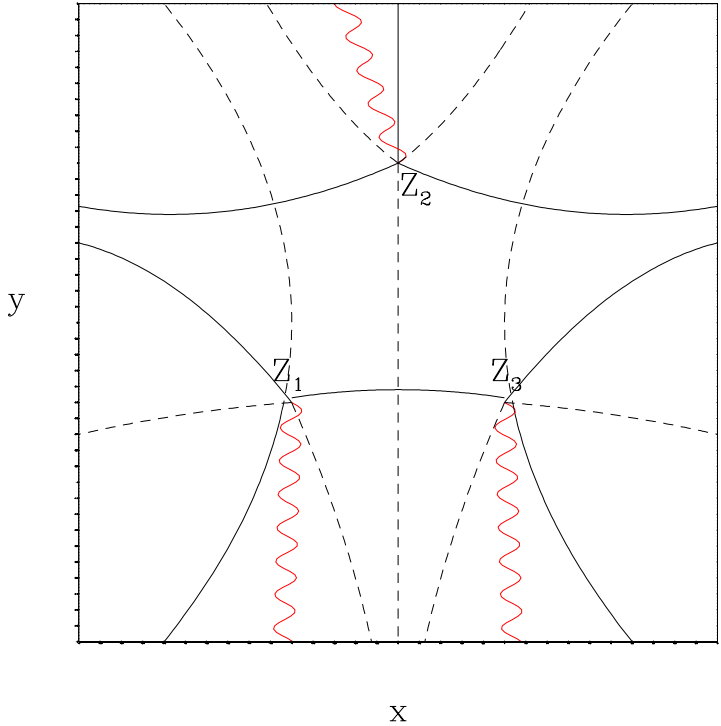}
\caption{Stokes plot for the bound state problem, $Q = E + (iz)^3$.}
\label{z3}
\end{figure}

The Stokes diagram is shown in Fig. \ref{z3}, with three singular 
points located at 
 $E^{1/3}e^{i\pi/2}$, $E^{1/3}e^{-i\pi/6}$, and $E^{1/3}e^{-i5\pi/6}$. 
 Subdominant regions include the positive and negative real axis for 
$|x| \rightarrow \infty$. We carry out 
the continuation in the upper half complex plane in order to take account of 
the singular point at $Z_2$.

In order to do the connections, we need the expressions 
$[k,l] = e^{\int_{z_k}^{z_l}i\sqrt{Q(z)}dz}$.  Note that the 
sheet of $i\sqrt{Q(z)}$ is defined by the cut locations, with 
the initial sheet determined by the fact that $(z_3,z)$ is 
subdominant for $x \rightarrow +\infty$ , meaning that
 $i\sqrt{Q}(z) = -real$ in this domain.

Carrying out the integrals then gives 
\begin{eqnarray}
[1,2] = e^{\sqrt{3}W/2}e^{-iW/2},\hspace{5mm}  [1,3] = e^{-iW},\hspace{5mm} 
 [2,3] = e^{-\sqrt{3}W/2}e^{-iW/2}
\end{eqnarray}
where $W = E^{5/6}cos(\pi/6)\int_{-1}^1\sqrt{1 - u^3}du$, 
and $\int_{-1}^1\sqrt{1 - u^3}du = 1.68262$.

Begin for positive real $x$ with
 $\psi(z) = (Z_3,z)_s$,
and continue to large negative $x$, giving a subdominant and a 
dominant term, which we set to zero, giving  
\begin{eqnarray}
0 = e^{\sqrt{3}W}(1 + S^2) + 2S^2cos(W) + e^{-\sqrt{3}W}S^2
\label{subz3}
\end{eqnarray}
leaving 
\begin{eqnarray}
 \psi(z) = -(Z_1,z)_s(Se^{-\sqrt{3}W} + Se^{iW}),
\label{leftz3}
\end{eqnarray}
very similar to the equations found in section \ref{sz4}, since 
the Stokes structure in the upper half plane is the same.  In this case 
the differential equation is not real, 
but choosing symmetric integration paths 
from $x = 0$ asymptotically along the Stokes lines to the right and to the 
left, and using the symmetry of $Q(z)$ we again conclude that the phases 
of the solutions for large positive $x$ and large negative $x$ are 
equal within a sign. 

The 
cut locations give the fact that whereas $Q^{1/4} = e^{i\pi/4}$ for large 
positive $x$,  $Q^{1/4} = e^{i3\pi/4}$ for large negative $x$, so in fact 
choosing the eigenfunction 
to be real for $x \rightarrow +\infty$ and requiring that it be real  
for large negative $x$ gives  
 from Eq. \ref{leftz3} $cos(W) + e^{-\sqrt{3}W} = 0$. Again in this case the 
asymptotic value of the Stokes constant is given by $S = i + O(\epsilon^2)$
with $\epsilon = e^{-\sqrt{3}W}$.

 The first few  energy levels for $N = 3$, with WKB
 and numerical values given by Bender,
 and values from this Phase Integral analysis 
are given in Table III.
The WKB ground state energy has an error of 6 percent, 
and the Phase Integral value an error of 0.6 percent.

\begin{eqnarray}
{\bf Table \hspace{1mm} III. \hspace{2mm} Energy \hspace{2mm}
 Levels  \hspace{2mm} Q = E - iz^3} \nonumber \\
\nonumber
\left( \begin{array}{c|c|c|c|c}
\hline
n & E_{exact}  & E_{wkb} & cos(W) & E_{PI} \\
\hline
0 & 1.1562 & 1.09427 &  -6.5834\times 10^{-2} & 1.1496\\
1 & 4.1092 & 4.08949 &  -2.8533\times 10^{-4} & 4.0892\\
2 & 7.5621 & 7.54898 &  -1.2366\times 10^{-6} & 7.54898\\
3 & 11.3143 & 11.3043 & -5.3598\times 10^{-9} & 11.3043\\
4 & 15.2916 & 15.2832 &  -2.3228\times 10^{-11} & 15.2832
\end{array}\right)
\label{ez3}
\nonumber
\end{eqnarray}

\section{Potentials \label{pot}}
\begin{figure}
\noindent
\includegraphics[scale=.5]{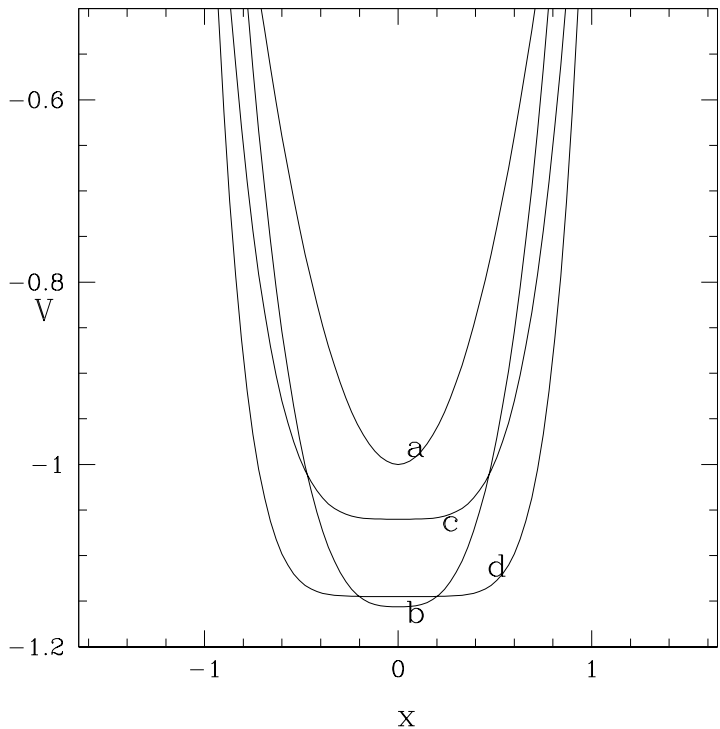}
\caption{Potentials for the four bound state cases discussed, the harmonic 
oscillator associated with the Weber equation, $V = x^2 - E$,  (a),
 the non-Hermitian potential
$Q = E + (iz)^3$  (b), and 
the Hermitian anharmonic oscillators $V = x^4 - E$ (c), and $V = x^6 - E$ (d).
The exact ground state energies were used for each plot.  }
\label{potential}
\end{figure}

The four potentials for the bound state cases are plotted in 
Fig. \ref{potential}.
The harmonic oscillator potential (a) $V(x) = z^2 - E$ and the two
anharmonic oscillator potentials (c) $V(x) = z^4 - E$ and (d) 
$V(x) = z^6 - E$ are real on
the real axis, $z = x$.  The potential associated with 
$Q = E + (iz)^3$ (b) is real giving subdominant solutions along the lines 
$y = -|x|/\sqrt{3}$, with $V(x) = |x|^3(\sqrt{3} - 1/3) - E$.
The energies used in the plot are the ground state values. 
It is seen that the degree of distortion from the harmonic 
oscillator potential shape is inceasingly larger for the 
$(iz)^3$ and the $z^4$ and $z^6$ cases, associated with the larger number 
of singularities in $Q(z)$ in addition to the two turning 
point singularities. We see that the error in the WKB energy levels 
increases with the deviation from the harmonic oscillator potential 
shape, along with a corresponding improvement of the Phase Integral 
evaluation over the WKB value. 

\section{Conclusion \label{con}}
A proper use of 
Phase Integral methods can improve the eigenvalue determination for 
bound states 
significantly compared to a simple WKB evaluation. This improvement increases 
along with the increasing deviation of the potential shape from that 
of a harmonic oscillator.  For all 
potentials possessing zeros or singularities in the complex 
plane in addition to the principal turning points, a small parameter 
is defined by $exp{\int(i\sqrt{Q})dz}$, with the integration taken
 from them to the principle turning points,
 allowing a perturbation expansion.  No such parameter exists for 
the Weber equation or the Budden problem.  However in these cases the 
Stokes constants can be calculated analytically.   It is remarkable that 
the asymptotic value of the Stokes constant in each case is $S=i$, 
the value for an isolated first order zero, and that in the cases examined in 
perturbation theory the corrections to $S$ 
are second order or higher in the small parameter given by the separation 
of the singular points. We have had to make the simplifying assumption that 
the Stokes constants are all equal, undoubtedly not true to higher order.  
The two complex equations resulting from the vanishing of the dominant 
solution and the symmetry or anti-symmetry of the subdominant solution 
do not allow proceeding to higher order, additional information is needed. 
It is an open question whether such relations exist, and whether the resulting 
equations lead to a convergent series giving the exact bound state energy. 
Of course the bound state 
energies do not form an open set, so $S$ is not determined as an 
analytic function of $E$, only the values at the bound eigenstates are fixed.
However, we conjecture that it is a common feature that all 
Stokes constants associated with lines emerging from a simple zero 
approach $i$ for large separation between singularities.

It has not escaped the authors' attention that
 with present day computing the numerically correct eigenvalues 
are easily obtained, so this result is only of theoretical interest.

\textbf{Acknowledgement}
This work was partially supported by the U.S. Department of Energy
Grant DE-AC02-09CH11466. The author acknowledges useful exchanges with 
Cheng Tang and constant encouragement from his grandson Enrico.

\end{document}